*Article*

# Recommender Algorithm for Supporting Self-Management of CVD Risk Factors in an Adult Population at Home

Tatiana V. Afanasieva *,1, Pavel V. Platov[2] and Anastasia I. Medvedeva[1]

1. Department of Informatics, Plekhanov Russian University of Economics, 36, Stremyanny lane, Moscow, Russia; afanaseva.tv@rea.ru
2. Department of Information Systems, Ulyanovsk State Technical University, 32 Severny Venetz Street,
   Ulyanovsk, Russia
* Correspondence: afanaseva.tv@rea.ru

**Abstract:** One of the new trends in the development of recommendation algorithms is the dissemination of their capabilities to support the population in managing their health**.** This article focuses on the problem of improving the effectiveness of cardiovascular diseases (CVD) prevention, since CVD is the leading cause of death worldwide. To address this issue, a knowledge-based recommendation algorithm was proposed to support self-management of CVD risk factors in adults at home. The proposed algorithm is based on the original multidimensional recommendation model and on a new user profile model, which includes predictive assessments of CVD health in addition to its current ones as outlined in official guidelines. The main feature of the proposed algorithm is the combination of rule-based logic with the capabilities of a large language model in generating human-like text for explanatory component of multidimensional recommendation. The verification and evaluation of the proposed algorithm showed the usefulness of the proposed recommendation algorithm for supporting adults in self-management of their CVD risk factors at home. As follows from the comparison with similar knowledge-based recommendation algorithms, the proposed algorithm evaluates a larger number of CVD risk factors and has a greater information and semantic capacity of the generated recommendations..

**Keywords**: knowledge-based recommender, multidimensional recommendation; multifactorial user profile; rules; large language model

## 1. Introduction

Recommendation systems as applied AI area that implement intelligence technologies have become commonplace in modern online e-commerce and e-services that have accumulated big data. A large number of research is conducted in domain of product and movie recommendations, whereas opportunities for recommender systems in health domain is not study enough.

Healthcare as a new area for using the capabilities of recommender systems has different directions and specific features, in particular, safety, evidence-based, high-risk decisions, face-to-face communications and poor formalization. The need to solve the problem of managing health indicators within the framework of the principles of personalized medicine using intelligent systems is emphasized in the study [1]. One emerging area of healthcare where the power of recommendation systems can have a significant positive impact is disease prevention, particularly cardiovascular disease. Cardiovascular diseases (CVD) are non-communicable diseases that are chronic in nature. A feature of the development of such diseases is a long asymptomatic period, which can lead to cardiovascular events requiring urgent medical care for the patient. The relevance of CVD prevention is associated with increasing life expectancy and due to the significant contribution of CVD to the total mortality of the working-age population. Therefore, early detection and prevention of CVD, including management risk factors, is an important focus of preventive services in all countries. To solve this problem, scientific research is being carried out [2], programs are being developed [3], algorithms [4], models [5] and clinical guidelines for the prevention of CVD [6, 7], intended for specialists of medical organizations. The recommendations of the European Society of Cardiology in 2021 noted for the first time that the task of preventing CVD must be solved at two levels: at the individual level, the goal of which is to personalize approaches as much as possible, and at the population level, which should be based on population mechanisms for controlling risk factors. Systematization of the main problems in effective strategies for the prevention of CVD at the individual level is discussed in the article [8]: a) lack of knowledge among the population about signs of diseases, risk factors and individual risk; (b) lack of digital tools for disease prevention; (c) insufficient physician time to provide detailed patient-centered prevention recommendations during clinical visits; (d) targeting prevention at high-risk individuals, whereas most acute cardiovascular events occur in low- or moderate-risk individuals.

According to official guidelines [6, 7], the main direction of CVD prevention is the management of risk factors not only in medical organizations, but at home too, as CV risk monitoring and assessment should be carried out regularly. CVD prevention considers non-modifiable and modifiable CV risk factors. The latter include behavioral risk factors which include unhealthy diet, tobacco smoking, insufficient physical activity, poor quality of sleep, regular stress, and biological risk factors such as high blood pressure, obesity, bad cholesterol and glucose levels.

There is challenge in self-manage of CV risk factors at home as they are characterized by diversity and heterogeneity, their indicators are presented on different scales, assessment of which depends on gender, age, non-modifiable CV risk factors and clinical conditions [9].

One of the promising ways to solve these problems is the creation of systems to support self-monitoring of CVD risk factors for the adult population at home the main purpose of which will be the creation of personalized recommendations for the CVD prevention.

The development of digital technologies has enabled the growth of digital prevention systems [10, 11], among which health recommendation systems (HRS) focused on self-management at home and promoting healthy lifestyles are promising for disease prevention [12–18]. HRS is a specialized recommendation system that, based on a user's profile (model), generates a list of recommendation items using information sources related to healthcare and health lifestyle. The main component of HRS is recommendation algorithm providing selecting recommended items, in which the selection conditions are determined by user profile, or by attributes of recommended items or based on domain knowledge [19-21]. As noted in the review [13], the most common HRSs are knowledge-based systems, where knowledge is represented in the form of rules. The study [9] demonstrated that a knowledge-based HRS has the potential to accurately recommend educational materials to chronic disease patients. Knowledge-based recommendation algorithms for self-manage of CVD risk factors at home are the most adequate too, as CVD prevention is a high-risk domain of healthcare in which precise and timely recommendations based on knowledge can prevent or delay fatal CV events.

At the same time incorrect recommendations obtained using another type of recommendation algorithm can have irreversible negative consequences for a person [19]. To avoid such situations, obtained recommendation should be based on the principles of evidence-based medicine, multifactor risk model and the knowledge outlined in official guidelines [6,7].

The goal of this study is to create a knowledge-based recommendation algorithm for supporting self-management of CV risk factors in an adult population at home, which will generate individual recommendations based on 2021 ESC Guidelines on cardiovascular disease prevention in clinical practice [6]. Utilizing knowledge from official guidelines for CVD prevention will ensure the completeness of the recommendations created, their safety and precision, which is critical when supporting persons outside the clinical setting. Let us note the global and long-term purpose of developing that algorithm is to reduce mortality from CVD, which is often caused by person's ignorance of their risk factors, lack of knowledge how to control them, and what conditions of their CV system require medical care, including urgent one.

Another motivation for developing knowledge-based recommendation algorithm for supporting self-management of CV risk factors in an adult population at home is the limitations and the gaps in research in this domain, namely:

1. The set of risk factors used is incomplete. Recommender algorithms utilize few CVD risk factors that characterize a person's lifestyle, in particular insufficient physical activity and/or unhealthy diet [13, 22, 23]. While for the prevention of CVD, it is important to make recommendations considering more range of risk factors, concerning behavioral, biological and non-modifiable ones [6,7].

2. Limitations of the user profile model. Recommender algorithms create recommendations based on the user's current profile and do not consider predictve assessments of the CVD development [13], Although, there is clinically approved family of prediction algorithms and tools to estimate 10-year risk of CVDs [4].

3. Lack of a formal model of the recommendation. Requirements for the recommendations being created and the formal model of the recommendations are not considered in the HRS studies, perhaps due to the fact that the recommendations are presented as a list of recommendations items of the same dimension of meaning, of what is recommended to be done, for example, "Read an online article about the symptoms of stress and good practices for stress management'" [24] or "Forget about sugar" [23]. However, during a preventive consultation, the physician gives the person professional advice of different dimensions of meaning, containing goals, ways to reduce the person's risks of cardiovascular disease, and explains why it is beneficial.

To overcome these limitations and fill the gaps in research of recommendation algorithms, this article presents knowledge-based recommender algorithm, called CVD-Risk-Prevent, for supporting self-management of multifactorial person's CV risk at home. The novelty of the proposed algorithm lies in combining rule-based models with large language models, namely ChatGPT, to generate multidimensional recommendations, which include recommendation items, outlined in official guidelines [6,7].

The contribution of the study is defined in more detail as follows:

In order to solve the problem of incomplete recommendations regarding the number and type of risk factors the multifactorial CVD risk model is presented based on the set of risk factors described in official guidelines [6,7]. By CVD risk factor we mean a characteristic of the user's condition that may be the cause of the development of CVD or an acute CV event, for example, myocardial infarction. Therefore, the model includes modifiable and non-modifiable CV risk factors, as well as an additional risk factor in the form of angina symptoms leading to a significant deterioration in well-being.

A user profile is proposed containing descriptive assessments of CVD risk factors and predictive ones of 10-year risk of CVD events. Predictive assessments are crucial for appropriate CVD preventive goal settings, and descriptive assessments inform person about their current risk factors and how to manage them.

The formal model of recommendation is described as a structure with 4 dimensions of meaning similar to the content of CVD preventive recommendation given by the physician. Each dimension of meaning is associated with one type of recommendation items, that are targeted, informational, explanatory and planned. Thus, multidimensional recommendations for any person have the same four-dimensional structure but with individual content, determined by the user profile.

Based on above models, CVD-Risk-Prevent algorithm was developed to support self-management of CV risk factors in adults at home. The algorithm uses knowledge representation in the form of rules, since the rules are clear, deterministic, transparent and scalable. The logic of the algorithm is based on rules of three types: rules for identifying individual CV risk factors, rules for descriptive and predictive assessment of CVD risk, and rules for creating multidimensional recommendations. The peculiarity of the algorithm for creating recommendations is that a large language model, ChatGPT (Generative Pre-trained Transformer), is also used to obtain the explanatory items of the recommendation.

The rest of the article is organized as follows. The second section provides a brief overview of the field of recommendation algorithms and recommender systems for person' support. A model of multidimensional recommendations based on a formal representation of multiple CVD risk factors, including 4 types of recommendation items, is presented in Section 3. The fourth section describes multifactorial user's CVD risk model. Section 5 presents the proposed recommendation algorithm, which we called CVD-Risk-Prevent, for supporting self-management of CV risk factors in an adult at home. Using ChatGPT for explanatory items creation is described and discussed in Section 6. In the Section 7 verification and evaluation of CVD-Risk-Prevent algorithm are provided. Discussion and conclusions are presented in the last section of the article.

## 2. Literature Review

In reviews [13, 14] types of recommendation items used in HRS are given. These include educational materials on diseases, diet, physical activity, motivation, medications, recommendations about clinics or doctors. In the study [15] provides an overview of recommendation algorithms, such as: collaborative filtering, content filtering, as well as knowledge-based filtering, including those obtained by machine learning methods (classification, clustering, decision tree methods).

To operate the collaborative filtering algorithm, the following input data are used: a set of users N, a set of recommendation items M, and a table of user ratings, representing the users' preferences towards the recommendation items [25]. The idea of collaborative filtering is based on the hypothesis: "If patients have similar diseases or health conditions, then they will have similar treatments, health services and lifestyle recommendations". The disadvantages of this scenario for the prevention of CVD include the unscientific nature of the recommendations, the need to have representative data on a variety of patients and the lack of relevance of the recommendations for a particular patient due to the individual characteristics of his CV health. The review [18] describes the various challenges present in current recommender systems. In particular, cold start, the scalability problem, data sparsity, synergy problem, fake ratings and fake user profiles for artificially increasing the rating of recommendation points. Among the promising areas of recommendation systems, the authors note research in the field of knowledge-based recommender systems.

Content-based recommendation algorithms also have as input a set of users N, a set of recommendation items M, and a rating table of recommendation items. The idea of the recommendation algorithm includes searching for similar of recommendation items, and not similar users as in the collaborative filtering algorithm [26]. In relation to supporting self-management of CVD risk factors, the limitations of scenarios for creating meaningful recommendations are redundancy, irrelevance of recommendations and the need to have a rating table, which is not always possible [16]. The unresolved problem of availability of high-quality data, the problem of sparseness of the rating table and "cold start" led to the development of systems for creating recommendations that use knowledge, such systems are called knowledge-based recommendation systems.

Knowledge-based recommendation algorithms take as input the data of one user X and, using a set of pre-generated rules, associate the user's profile with a set of recommendation items H⊂M. Unlike recommender systems that create recommendations for similar recommendation items or similar patients, knowledge-based recommender systems can create recommendations that consider the domain knowledge, limitations, health conditions, and preferences of a particular patient. The use of inference rules, machine learning techniques, mathematical models and explicit knowledge is a key feature of knowledge-based recommendation scenarios [17]. The sources of recommendation items in such algorithms are expert knowledge, knowledge extracted from official medical guidelines and knowledge about the person's health status. Authors of review [19] emphasize that, unlike other recommender algorithms, the fields of application of knowledge-based recommendation are those in which irrelevant recommendations can cause serious negative consequences. In the paper [27] user needs for explanation of recommendation received by recommender systems are studied including knowledge-based recommendation. That is why, knowledge-based recommendation algorithms have become more widespread in HRS [13] and promising in explainable AI systems for the management and monitoring chronic patients aimed at fostering a healthy lifestyle [9,28].

A study [23] focuses to create a model for recommending challenges that positively impact the health of individuals. The model includes characterizing the user, challenges, and knowledge-based algorithm for generating appropriate suggestions. The user profile is characterized by a total of 24 features for persons without medical conditions and considers physical, mental, and nutritional aspects. Article [24] notes the limitations of knowledge-based recommendation algorithms of the range of lifestyle risk factors used, namely physical activity and diet. They proposed knowledge-based recommender algorithm expanding multiple lifestyle risk factors from a multidimensional viewpoint. The recommendations received were intended to support identifying appropriate coaching plans for stress management. The authors of the work [1] propose an approach in which, based on a heterogeneous semantic network and rules, recommendations are created for lifestyle changes to reduce the risk of CVD. In this research the set of lifestyle risk factors was extended with non-modified ones. The limitations of knowledge-based HRS, according to modern publications, are the incompleteness of medical knowledge, the lack of ability to monitor and analyze trends in changes in the person's health status, and the lack of formal descriptions of the recommendation models used [29].

In a recently published article [30] the author focuses on presenting the logical foundations of knowledge-based recommender systems, a specific branch of AI research, where recommendations are the result of problem solving on an explicit knowledge base that links user properties and preferences to categories and /or items to recommend. The review [31] provides a variety of metrics used to evaluate recommendation systems in the field of e-commerce and notes that one of the ways to evaluate recommendation algorithms is to evaluate the utility of recommendations received by end users. The authors emphasize the need for research in a new direction that will effectively assess user satisfaction with the recommendations received. In the Handbook of Recommendation Systems [32], the authors argue that utility is the value that users receive when they are recommended, that is, utility can be measured by evaluating recommendations after they are provided to the user. Integration of rule-based and natural language processing (NLP) approaches to create knowledge-based recommendation on personalized educational materials for chronic disease patients in China was proposed in the study [9]. The patient profile was presented using ontology, and then user vector was generated by the 80 SWRL rules, previously defined. The ontology structure includes patient characteristics in 5 areas: chronic diseases, lifestyle and biological risk factors, demographic features and medication use. NLP technology was applied to obtain text vectors based on the keywords of each educational document. Inner product as a tool to indicate the relevance between patient data and educational materials was proposed. Recent advancements in natural language processing have introduced Large Language Models (LLMs) that exhibit remarkable capabilities in understanding and generating human-like text [33]. As noted in [19] the implementation of LLM in knowledge-based recommender systems is beginning to play an important role in the context of various tasks, including the creation of explanations in specific recommendation contexts. These models have the big potential in cardiology improving medical diagnosis and decision support by analyzing vast amounts of medical data and provide educational materials for patients [34]. Based on the results of the analysis, the following conclusions can be drawn. Knowledge-based HRS is an innovative tool for digital CVD prevention, in which recommendation algorithms to support adults at home are promising and under active development. However, to the best of our knowledge, no studies to date have formally concentrated on a knowledge-based recommendation algorithms for supporting adults at home in self-management of the set of CV risk factors in outlined in official guidelines [6,7].

## 3. Formal Model of Multidimensional Recommendation

Recommendation items are the main components of knowledge-based recommender algorithms. When developing a recommender algorithm for supporting self-management CV risk factors at home, it is necessary to consider that it should be intended both for individuals without CV risk factors, for generally healthy individuals and for individuals with existing CVD, which may cause acute CV events requiring urgent medical attention.

Analysis of HRS research [12-17], the problems in effective strategies for the prevention of CVD at the individual level [8] and the process of consulting persons on the prevention of CVD, carried out by preventive doctors, allows us to determine the following requirements for the recommendation content:

1. Usefulness and safety. The items in the recommendation used should be specified in current clinical guidelines for the prevention of CVD. The recommendation for supporting self-management at home should be within the framework of the principles of non-pharmacological intervention.

2. Completeness. The recommendations should include all CVD risk factors and their assessments, as well as predictive assessments, outlined in current clinical guidelines for the prevention of CVD. At the same time the set of CVD risk factors should include symptoms of angina pectoris as the main predictor of coronary heart disease.

3. Accuracy and Individual focus. Recommendation items must be correct and free of errors. The text of recommendation should provide information about person's CV risk factors only if they are identified. For the persons without CV risk factors the text of recommendation should present general information supporting their healthy lifestyle.

4. Multidimensionality. The text of recommendation should combine items of different dimensions of meaning containing goals, ways to manage the person's CV risks, and explanations why it is beneficial.

5. Planning. An action plan is a key component of the recommendations. The recommendation items in the action plan will provide the user with information about what steps to take for self-management their CV risk factors.

6. Timeliness. The recommendations must be supplemented with information about the degree of urgency of seeking medical help depending on the assessment of CVD risks.

7. Explainability. The recommendation items should not only provide information, but also explain its importance for the management of the person's CV risks.

8. Motivation. The recommendation should support the person to take the first step in direction of care about their CV health and support the formation of the skill of self-managing their CV risk factors.

Based on the above-defined requirements, a recommendation model will be understood as a text, the content of which combines recommendation items of different dimensions of meaning and contains the individual characteristics of the CV health of each person, grouped by the dimensions. Then we propose a formal model of multidimensional recommendations for supporting self-management CV risk factors in adults at home in the form of a set of recommendation components of different dimensions, namely: targeted $Es$, informational $Inf$, explanatory $Expl$ and recommendation component in the form of an action $Plan$:

$$R = \{Es, Inf, Expl, Plan\}, \tag{1}$$
$$Es = \{EsG, EsR\}, \tag{2}$$
$$EsR = \{EsR(1), EsR(2), \ldots EsR(N)\}, \tag{3}$$
$$Inf = \{Inf(1), Inf(2), \ldots, Inf(N)\}, \tag{4}$$
$$Expl = \{Expl(1), Expl(2), \ldots, Expl(N)\}, \tag{5}$$
$$Plan = \{p(1), p(2), \ldots, p(N)\}, \tag{6}$$

where N is the number of person's CV risk factors.

The targeted recommendation component includes goals at the strategic level $EsG$ and tactical level $EsR$ containing recommender values and trends for each CV risk factor that requires correction for the next time. The $Es$ component answers the question "What is recommended to be done?" and contains information about the degree of urgency and type of medical care too. Below the example of target component is presented.

EsG: "CVD prevention is your main goal. Taking into account your risk factors, you are advised to consult a therapist to receive medical care and possibly additional instrumental examination". EsR: "It is necessary to lower blood pressure, reduce salt intake and stop smoking. Target values for your age: blood pressure no more than 130 mmHg".

The informational component $Inf$ contains descriptive and predictive assessments the state of CV health, for instance, "Your total CV risk over a ten-year period is classified as moderate. Your CV risk factors are: high blood pressure, unhealthy diet, and smoking".

The explanatory component $Expl$ answers the question "Why is this recommended?". The examples of explanatory items are:

1." High blood pressure is the main risk factor for CVD, as it increases the load on the heart. The more salt a person consumes, the higher the risk of premature death from any cause. For an average 50-year-old woman, excess salt loses 1.5 years of life, and for a man – 2.3 years of life and significantly increases the risk of arterial hypertension".

2. "Stop smoking, because smoking is one of the main factors in the high risk of death from cardiovascular events and other chronic diseases. Quitting smoking can reduce this CVD risk by half. Those who do not smoke have an average life expectancy of 10 years longer. The risk of CVD in smokers under 50 years of age is five times higher than in non-smokers; after just a few puffs, the heart begins to work hard - carbon monoxide from tobacco smoke sharply reduces the concentration of oxygen in the blood, and it has to be distilled in large volumes to prevent the body from feel oxygen deficiency".

An individual plan of effective action $Plan$ answers the question "How to implement the recommendations?" and is focused on achieving the goals defined in the $Es$ component in brief or in detail. It may include recommendation items on lifestyle, optimization and correction of nutrition, relevant physical activity, a schedule for monitoring and screening studies, consultations with medical specialists, including emergency ones:

Plane: "Did you know that 90% of people who promise themselves to become better do not keep this promise because most people do not understand the process of creating sustainable results, so they never take the first step. An effective action plan will give you the start and the driving force that will lead you to improve the quality and duration of your active life. Your plan for effective action:

A. Consultation with a therapist as soon as possible. To lower blood pressure, drug therapy is required, which will be determined by your doctor.

B. quit smoking.

C. Gradually increase your physical activity, starting with 30 minutes of walking a day".

The proposed multidimensional recommendation model (1) has a structure that is as close as possible to the structure of clinical recommendations for the prevention of CVD [6] regarding recommendations for management health indicators, CV risk factors, goals, and a plan to achieve goals without recommendations for treatment. The key difference between the proposed recommendation model and those used in other knowledge-based recommendation algorithms is its structure and semantic multidimensionality, corresponding to the logic of counseling on the management of CVD risk factors.

4. Multifactorial user's CVD risk model

As a result of the analysis of guidelines for CVD prevention [6,7], the following conclusion was made, the user profile should be based on a modern model of CV risk factors outlined in official recommendations, among which the ESC 2021 Guideline [6] and its version [7] were selected in this study. By a CV risk factor, we mean a characteristic of the user's condition, which may be the cause of the development of CVD or an acute CV event, for example, myocardial infarction. Therefore, the model should include modifiable, non-modifiable and specific CVD risk factors described in [6], as well as an additional risk factor in the form of angina pectoris symptoms leading to a significant deterioration in well-being. Taking in consideration the multifactorial structure of CVD risk factors, it is important to determine its formal model before defining a user profile on this basis. Let's define a multifactorial model of CV risk factors in the form of the following expressions:

$$Z = \{S, NonModF, Biol, Beh\}, \quad (7)$$
$$NonModF = \{CVD, KD, DM2, GEN\} \quad (8)$$
$$Biol = \{OBS, CH, NHL, SBP, GL\} \quad (9)$$
$$Beh = \{Sm, Fa, Diet\}. \quad (10)$$

The variable designation used in expressions (7-10) is presented in Table 1, where the sign of the CV risk factor is coded by value "1", and its absence - by value "0".

In the formal model $Z$, we use 13 CVD risk factors grouped by their types. In accordance with [6] when defining person's CVD risk the class of person and prediction assessment should be done.

Table 1. Designation of CV risk factors. If there is a risk factor, its value will be set to "1", otherwise "0"

| Factor No. | CV risk factor designation and description |
| --- | --- |
| Type 1: Symptoms | |
| 1 | $S$ — a sign of symptoms of angina pectoris with a significant deterioration in health: a) paroxysmal pain, pressing, squeezing, burning, in the chest in the left region, the lower part of the sternum along with the upper abdomen, even with weak or moderate intensity; b) lack of air, shortness of breath, irregular heartbeat, severe weakness, cold sweat, nausea; c) increased frequency of attacks angina for which nitroglycerin does not respond within 15 minutes |
| Type 2: Non-modifiable factors (NonModF) | |
| 2 | $CVD$ — a sign that the person has documented one of the CVD (arterial hypertension, coronary heart disease, atherosclerosis, myocardial infarction, stroke, chronic heart failure, cardiac arrhythmia, cerebrovascular disease) |
| 3 | $KD$ — a sign of chronic kidney disease |
| 4 | $DM2$ — a sign of type 2 diabetes |
| 5 | $GEN$ — sign of family history of early CV diseases in close relatives: in men under the age of 55 years and in women under 65 years |
| Type 3: Modifiable biological factors (Biol) | |
| 6 | $OBS$ — a sign of obesity if Body Mass Index is more than 24 kg/m |

| 7  | CH – a sign of increased total cholesterol, if it is more than 5 mmol/l |
| --- | --- |
| 8  | NHL – a sign of increased levels of non-high-density lipoprotein cholesterol, if it is more than 3 Mmol/l |
| 9  | SBP – a sign of increased systolic blood pressure, if it is more than 140/90 mmHg |
| 10 | GL – a sign of increased glucose levels in the blood serum, if it is more than 7 mmol/l |
| Type 4: Modifiable behavioral factors of lifestyle (Beh) | |
| 11 | Sm– a sign of tobacco smoking |
| 12 | Fa – a sign of insufficient physical activity, no more than 2 hours of household activity or moderate-intensity aerobic physical activity per week |
| 13 | Diet – a sign of unhealthy food abuse |

Taking in account these recommendations the model of user profile is presented as
$$UP = \{Factor, C, Class\}. \qquad (11)$$
Here $Factor$ is a binary vector of descriptive assessments of individual CVD risk factors based on the model $Z$, $C$ denotes the predictive assessment of total CVD risk in 10-year perspective, commonly used in clinic practice by application of SCORE algorithms [4,6,7], and $Class$ is a binary vector which components describe person belonging to one of the risk factor types presented in Table 1. Assigning individuals to different classes is necessary to determine the recommendation items containing the strategic goals of CV prevention, and the urgency of providing medical care. The use of not only descriptive but also predictive estimates is a key advantage of the proposed user profile model for creating recommendations with greater information capacity. To determine the user profile $UP$ we use the set of CVD health indicators in the form
$$X = \{x_j, x_j \in \mathbb{R}, \ x_j \geq 0, \forall \ x_j \in [\min\_x_j, \max\_x_j], j = 1,2\ldots 17\}. \qquad (12)$$
Table 2 shows the set of CV health indicators, grouped by types of CV risk factors. Questionnaires, texts from electronic medical records, and data from digital medical gadgets can be used to gather user CVD health indicators [12-18].

Table 2. Indicators of a person's CV health. Here 1 denotes "Yes", 0 – "No"

| X | Indicator | X | Indicator |
| --- | --- | --- | --- |
| Basic indicators | | Modifiable biological indicators | |
| X(1) | Gender: male (1)/female (0) | X(10) | Total cholesterol (mmol/l) |
| X(2) | Age (years) | X(11) | Non-high-density lipoprotein cholesterol (mmol/l) |
| X(3) | Height (cm) | X(12) | Systolic blood pressure (mmHg) |
| X(4) | Weight (kg) | X(13) | Glucose level (mmol/l) |
| Non-modifiable indicators | | Modifiable lifestyle indicators | |
| X(5) | Family history of CVD (1/0) | X(14) | Physical inactivity (1/0) |
| X(6) | Presence of CVD (1/0) | X(15) | Smoking (1/0) |
| X(7) | Chronic kidney disease (1/0) | X(16) | Unhealthy diet (1/0)) |
| X(8) | History of CV events (1/0) | Symptoms of CVD or CV event | |
| X(9) | Diabetes mellitus type 2 (1/0) | X(17) | Symptoms of angina pectoris, with a significant deterioration in health (1/0) |

## 5. Recommendation Algorithm

We propose a knowledge-based recommendation algorithm, namely CVD-Risk-Prevent, for supporting adults in self-management their CVD risk factors at home, based on official guideline [6]. The main user of the recommendation algorithm is a person interested in consciously managing their health to increase the duration of an active and full life. Considering that such individuals may have various CV risk factors, including CVD, they are forced to regularly visit their doctor, preventive care offices, and/or perform self-monitoring and control of basic CV health indicators. The key feature and contribution of this algorithm is the formal description of the transformation of one data structure in the form of a set of person's CV health indicators X (see Table 2) into another data structure presented in the form of multidimensional recommendations R (see expressions 1-6), using formal model of user profile UP (11), a set of rules and LLM.

*5.1. Problem Statement*

Let us present the statement of the problem to be solved by the CVD-Risk-Prevent algorithm to support self-management of CV risk factors at home. We assume that we are given: a set of CV health indicators X for a person (see Table 2), a model of CV risk factors Z presented in Section 4, and a multidimensional model of recommendation R described in Section 3, includes the item sets of four different types. It is required for this person to generate recommendation R* that meet the requirements to recommendations, introduced in Section 3, that is, to construct a mapping

$$X \times Z \times R \to R^*, \quad (13)$$

where × is the sign of the Cartesian product, R* is the recommendation, containing items of each type $Es, Inf, Expl, Plan$ corresponding to model R (1) and user profile UP (11).

*5.2. The CVD-Risk-Prevent Algorithm*

The CVD-Risk-Prevent algorithm for each person includes the following stages:

Using CV health indicators X and model Z of CV risk factors creation a descriptive component for user profile UP, that is a vector $Factor$.

Determining a descriptive component $Class$ of user profile UP based on vector $Factor$.

Assessment of total CV risk over a 10-year period using the SCORE/SCORE2-OP models [4, 6] and calculation of predictive estimate C of user profile UP.

Mapping user profile UP with recommendation items $Es, Inf, Expl, Plan$, containing in the model R.

Let's consider these stages in more detail.

Stage 1. Using CV health indicators X and model Z of CV risk factors (7) creation a descriptive component for user profile UP, that is a vector $Factor$. If a person did not specify CV health indicator in X, this indicator is set to 0. Since the components in CV risk factor model Z are given in Table 1 by binary signs (1–yes, 0–no), we propose to use a binary vector $Factor$ which initial values of each component are set to 0:

$$Factor = (f(1), f(2), \ldots, f(13)), f(i) \in \{0,1\}, i = 1,2,\ldots,13 \quad (14)$$

here $i$ – CV risk factor number, presented in Table 1.

To determine values of the vector $Factor$, we use the rules based on calculations grouped by the types of CV risk factors. To calculate the signs for non-modifiable CV risk factors, we define the following computational rules:

$$f(1) = x(17), \ f(2) = x(6) \lor x(8), \ f(3) = x(7), \ f(4) = x(9), \ f(5) = x(5). \quad (15)$$

For modified behavioral factors, vector components are also calculated directly from CV health indicators using the expressions:

$$f(11) = x(15), \ f(12) = x(14), \ f(13) = x(16). \quad (16)$$

For a biological CV risk factor obesity $f6$, body mass index BMI is pre-calculated:

$$BMI = \begin{cases} 10^4 \cdot \frac{x(4)}{x(3)^2}, & if\ x(3) \neq 0, \\ 25, & if\ x(3) = 0, \end{cases}$$

and then the value of the $f6$ component is determined using expression corresponding [6,7]:

$$f(6) = \begin{cases} 0, & if\ BMI < 24, \\ 1, & if\ BMI \geq 24. \end{cases} \quad (17)$$

For biological CV risk factors, the components $f(7), f(8), f(9), f(10)$ of the $Factor$ vector will be determined using the rules for comparing values $x_q, q = 10,11,12,13$ with the $p_q$ thresholds specified in clinical guidelines [6,7] and presented in Table 1:

$$f(j) = \begin{cases} 0, & if\ x_q \leq p_q, \\ 1, & if\ x_q > p_q, \end{cases} j = 7,8,9,10; q = 10,11,12,13. \quad (18)$$

Note that if a person did not specify indicator the corresponding component of the $Factor$ vector will remain equal to 0.

Stage 2. Determining a descriptive component Class of user profile UP (11) based on vector Factor (14). Using types of CVD risk factors in multifactorial model Z (7) and vector Factor the CVD-Risk-Prevent algorithm classifies persons into 5 classes:

Class 1. Individuals are generally healthy and do not have CV risk factors or if a person did not specify CV health indicators. We will classify such persons as persons of class 1 without CV risk factors.
Class 2. Persons for whom modifiable behavioral risk factors for CV risk have been identified.
Class 3. Individuals for whom modifiable biological CV risk factors have been identified.
Class 4. Individuals for whom non-modifiable CV risk factors have been identified.
Class 5. Persons who have symptoms of angina associated with a sudden significant deterioration in health. We classify such persons as ones needed medical urgent care.

Elicitation of the person class is necessary to create a target setting for the person, which will first of all inform the person whether he/she needs medical care, its regularity and urgency. Considering the situation that a person can have different CV risk classes identified, we will simultaneously determine the person CV risk class in the form of a binary $Class$ vector with initial values zero of each component:

$$Class = \{Class(1), Class(2), Class(3), Class(4), Class(5)\}. \tag{19}$$

To evaluate the person's class, the following logical rules are used, which change the corresponding component of the Class vector to the value 1 analyzing the vector Factor (14):

$$Rule\ 1: IF\ \max_{i=1,2,\dots,13} f(i) = 0\ THEN\ Class(1) = 1, \tag{20}$$

$$Rule\ 2: IF\ \max_{i=11,12,13} f(i) = 1\ THEN\ Class(2) = 1, \tag{21}$$

$$Rule\ 3: IF\ \max_{i=6,7,\dots,10} f(i) = 1\ THEN\ Class(3) = 1, \tag{22}$$

$$Rule\ 4: IF\ \max_{i=2,3,4,5} f(i) = 1\ THEN\ Class(4) = 1, \tag{23}$$

$$Rule\ 5: IF\ f(1) = 1\ THEN\ Class(5) = 1. \tag{24}$$

According to the above rules and vector $Class$, the person can be assigned to one of 16 sub-classes. One sub-class will include persons without CV risk factors, and the remaining ones will include persons with various combinations of CV risk factors and clinical conditions.

Stage 3. Assessment of total CV risk over a 10-year period using the SCORE/SCORE2-OP models [4, 6, 7] and calculation of predictive estimate C of user profile UP (11). The SCORE (Systematic Coronary Risk Evaluation) model is designed to estimate (probably predict) the risk of fatal CVDs over a 10-year period, considered in guideline for CVD prevention [6]. To assess the total CV risk of a person using SCORE, the following are used: person's CV health profile (Table 2): x(1) - gender, x(2) - age, x(10) - total cholesterol, x(12) - systolic blood pressure, x(15) - sign of smoking. Subsequently, the SCORE2/SCORE2-OP model appeared presenting an expanded, updated prediction model to estimate 10-year risk of fatal and nonfatal cardiovascular diseases. Unlike the SCORE model, the SCORE2/SCORE2-OP model uses a person profile indicator of x(11) (non-high density lipoprotein cholesterol) instead of x(10). The updated SCORE2/SCORE2-OP model is applicable to individuals without prior CVD or diabetes aged 40–89 years in Europe.

The probability of the person's total CV risk in a ten-year perspective is calculated as CVrisk using the Rule 6, in which FUNC denotes the SCORE2-OR algorithm proposed in [4]:

$$Rule\ 6: IF\ Sco = 0\ THEN\ CVrisk = FUNC(x(1), x(2), x(11), x(12), x(15)) \tag{25}$$

$$here\ Sco = f(1) \lor f(2) \lor f(4) \lor f(5). \tag{26}$$

The logical expression (26) sets a constraint on the use of the SCORE2/SCORE2-OP algorithm outlined in guideline [6]. We consider SCORE2/SCORE2-OP algorithm as external function FUNC and obtain predictive assessment of person total CVD risk based on Rule 5 following guideline [6]:

$$C = \begin{cases} low, & if\ CVrisk < 2.5\% \\ moderate, & if\ 2.5\% \leq CVrisk < 5\% \\ high, & if\ 5\% \leq CVrisk < 10\% \\ very\ high, & if\ CVrisk \geq 10\%\ or\ (Sco = 1). \end{cases} \tag{27}$$

At the end of this stage all components of user profile UP (11) are defined and could be used to generate multidimensional recommendation for supporting self-management of person CV risk factors at home.

Stage 4. Mapping user profile UP with recommendation items. Please note that the contents of recommendation items $Es \in R, Inf \in R, Plan \in R$ were manually extracted from the guidelines [6,7], while the content of recommendation items. $Expl \in R$ was generated by LLM, namely ChatGPT--. Algorithm generate recommendation to a person basing on user profile components. For a person CV risk factors recommendation items associated with a non-zero value of the vector component $Factor$ (14) are generated:

$$EsR = \{EsR(i)\}, \tag{28}$$
$$Inf = \{Inf(i)\}, \tag{29}$$
$$Expl = \{Expl(i)\}, \tag{30}$$
$$Plan = \{p(i)\}, i = 1,2,\ldots,13. \tag{31}$$

To make the received text of the recommendations more diverse without violating their semantics an explanatory component $Expl$ of the recommendation were created using ChatGPT-4, since this model is able to generate human-like text [33] including the creation of explanations in specific contexts [19] in particular in cardiology [34].

The content of recommendation items of the EsG type which sets a strategic goal for the CVD prevention is determined by non-zero component of the Class vector and predictive estimate C. Therefore, to generate this type of recommendation items, the following rules are used, ranked in descending order of class number j:

$$Rule7j: IF\ Class(j) = 1\ THEN\ EsG(j), j = 1,2,\ldots,5. \tag{32}$$

Descending ordering of the $Class$ number is advisable because the larger the class number, the higher the risk of CVD in a person. For obtaining the tactical goals $EsR$, informational $Inf$, explanatory $Expl$ and planned $p$ components of user recommendation, the rules corresponding to non-zero values of the vector Factor (14) are used:

$$Rule\ 8i: IF\ f(i) = 1\ THEN\ EsR(i) \cup (C \cup Inf(i)) \cup Expl(i) \cup p(i), i = 1,2,\ldots,13. \tag{33}$$

In expression (33) the sign "∪" denotes the operation of concatenation of the items, and $i$ designates the risk factor number.

The ranking of recommendation items obtained using $Rule\ 8i$ (33) is carried out in 2 stages. Since each non-zero $f(i)$ corresponds to a single user Class (see formulas 21-24) at the first stage, the recommendation items in expression (33) are ordered in descending order of the user's Class numbers. Thus, at first, the user will be shown recommendation items concerning the most serious CV risk factors. Then, for user of classes 2 and 3, the recommendation items are additionally ordered in ascending order using the Utility function, as suggested in [35]:

$$Utility(h) = \sum_{i=6}^{13} f(i) \cdot S_i(h),$$

here h is a recommendation item, $f(i)$ denotes i-th component of Factor vector, $S_i(h)$ designates contribution of recommendation item h to i-th CVD risk factor. We assumed that the contribution value of the recommendation item y corresponds to the rank of the negative impact of CV risk factors on CVD and CV events considered in the Prospective Urban Rural Epidemiology (PURE) study [36]. The PURE study focused on examining the influence of CV risk factors on CVD and CV events in 155 722 people. As a result of this study, risk factors for CVD were ordered by their ranks, where the lower the rank, the greater the degree of its negative impact on this risk factor.

Figure 1 shows a framework of CVD-Risk-Prevent algorithm where the ovals show the stages of the algorithm, and the rectangles show the models and estimates.

The recommendation items that make up the recommendation are indicated by yellow polygons. Since the CVD-Risk-Prevent algorithm assumes the assessment of the predictive estimate of total CV risk using the SCORE2/SCORE2-OP model, to obtain the CV risk probability value, the medical calculator recommended by the EAPC (European Society of Cardiology) for CVD prevention [37] and the SCORE2/SCORE2-OP scale given in [4, 6] were used. To obtain initial data about person CV health, a questionnaire was developed, the questions of which correspond to CVD health indicators represented in Table 2. To assess the symptoms of angina pectoris, the Rose questionnaire was used [38].

## 6. Using ChatGPT for explanatory items creation

We decide to use ChatGPT-4 for generating only explanatory items $Expl$ of the recommendation for two reasons. Firstly, the explanation of the negative impact of the CVD risk factor is not sufficiently presented in the guidelines [6,7], and secondly, in our opinion, large language model can create unusual metaphors that could draw attention to the main content of the recommendation.

In the review [41], the authors emphasize that modern recommendation systems lack the explicability of recommendations, as well as knowledge about the open world to improve the effectiveness of recommendations. While large language models are promising for solving these problems, since LLM usually uses more than a billion parameters and is pre-trained on large volumes of text corpora containing up to trillions of tokens obtained from various Internet sources. With extensive knowledge of the open world, LLM can significantly improve the effectiveness of recommendations, in particular, in cardiology, was pointed in [34].

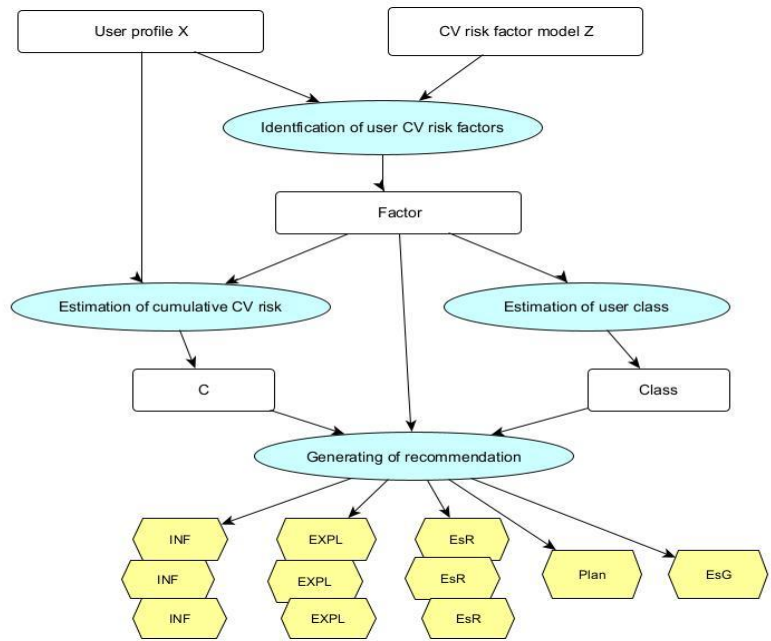

Figure 1. Framework of the proposed algorithm for the prevention of CVD

These new trends in the development of recommendation systems are especially relevant for knowledge-based recommendation systems designed to support the management of CVD risk factors at home. CVD-Risk-Prevent algorithm turns to ChatGPT with a prompt to explain why the identified CVD risk factors should be managed. We used the following general form of prompts to obtain ChatGPT responses:

$Response = ChatGPT(task, params, constraints)$,

where "task" is a text expressing the content of what we want to receive; "constraints" denote some restrictions for the response and "params" define the parameters that need to be taken into account, namely CVD risk factors. So, a prompt to ChatGPT contains three components as shown in Figure 2 where the blue symbols refer to LLM responses to the prompt.

| Please give the explanations why person need to control her/his CV risk factors such as high blood pressure – 160/90 mmHg, physical inactivity, smoking, unhealthy diet. The explanation must be understandable to the person, include only person CV risk factors and contain no more 4 propositions for each CVD risk factors.<br><br>Let's go over the person cardiovascular (CV) risk factors:<br><br>### 1. **High Blood Pressure (Systolic blood pressure – 160/90 mmHg)**<br>High blood pressure is like putting too much air in a balloon—it strains the walls and can cause them to burst. In the body, this can lead to damage in the arteries, heart, and other organs. It's often called the "silent killer" because it doesn't have symptoms but can lead to serious problems like heart attacks and stroke. Keeping blood pressure lower reduces the strain on your heart and the risk of heart disease.<br><br>### 2. **Physical Inactivity**<br>Being inactive is like leaving a car unused for a long time; eventually, it won't run as well. Regular physical activity keeps your heart and body running smoothly and can help lower high blood pressure, manage weight, and reduce stress on your heart. It can also improve your mood and energy levels.<br><br>### 3. **Smoking**<br>Smoking damages your blood vessels and can make them narrow, making it harder for blood to get around your body. Think of it as a traffic jam in your arteries. Stopping smoking can help to clear the traffic and reduce your risk of heart disease, as well as improve your overall health.<br><br>### 4. **Unhealthy Diet**<br>Eating foods high in fat, salt, and sugar is like using low-quality fuel in a car—it won't run properly and will break down sooner. A healthy diet is like high-quality fuel; it makes sure everything runs smoothly and reduces the risk of blockages in your arteries. It's good for your whole body, including your heart.<br><br>Addressing these risk factors involves lifestyle changes, such as getting more exercise, eating a healthier diet, and quitting smoking. These changes can significantly lower the risk of developing serious CV conditions, leading to a healthier heart and a longer life. It's like maintaining a car; you do it not just to prevent it from breaking down, but to make sure it runs well for a long time. | Please give the explanations why person need to control her/his CV risk factors such as moderate total CV risk on SCORE model, family history of early CV diseases, high blood pressure.<br>The explanation must be understandable to the person, include only person CV risk factors and contain no more 4 propositions for each CVD risk factors.<br><br>That's why now is the time to self-manage your cardiovascular risk factors:<br><br>### 1. ** Moderate Total CV risk **<br>You are at moderate risk for cardiovascular disease according to the SCORE model. This means you have a 1-4% chance of developing serious heart or vascular problems within the next 10 years. This is not the highest level of risk, but it is not the lowest, which means we need to pay attention to your health now to prevent possible problems in the future.<br><br>### 2. ** Family history**<br>Having cardiovascular disease in your immediate family increases your own risk. This means you should be more careful about prevention and get regular medical checkups to ensure any potential problems are caught and treated early.<br><br>### 3. ** Blood Pressure **<br>Please note that elevated blood pressure (BP) may be a sign of incipient hypertension. Systematic elevation of blood pressure is a major risk factor for heart disease, stroke and other serious health problems. As blood pressure rises, the load on the heart and blood vessels increases, and the heart begins to work harder.<br><br>Not smoking greatly helps reduce your risk. |

Figure 2. Two examples of the prompts and the explanatory item lists received as responses from ChatGPT

The first sentence of the prompt combines the task text and person CV risk factors, while restrictions on the reasoning for explanatory items are defined in the second sentence of the prompt. The prompt parameters "*params*" are a person's CVD risk factors, which are contained in the information component of his recommendation in the form of a list items identified in formulas (27, 33) as $(C \cup Inf(i))$.

Since ChatGPT is used for the first time to create explanatory items of recommendations for self-management of CV risk factors in adults at home, we formulated research question: Is the content of the generated explanations complete, non-redundant, and correct? When responding to a prompt containing one parameter, ChatGPT creates a semantically relevant and stylistically correct explanation of the specified risk factor within the specified restrictions. If the prompt contains a list of risk factors, the ChatGPT can create an explanation of them, united by one metaphor, as follows from the examples depicted in Figure 2. Interestingly, the ChatGPT additionally creates a generalizing explanatory element that leads at the end or beginning of the generated explanations as one could see in Figures 2 and 3 respectively.

To answer the research question, we also examined the ability of the used LLM to take into account additional parameters and logically insert them into the text of the explanatory items. The examples presented in Figure 3 illustrate the responses obtained on the same prompt as in examples presented in Figure 2 but with different parameter lists. Figure 3a shows the prompt parameters at the top, and below is the response text with which three generated responses received in prompts with a slightly expanded set of parameters will be compared. These three responses, together with the used parameters at the top, are presented respectively in Figures 3b, 3c and 3d. Changes in the texts of explanatory items with an expanded composition of prompt parameters are highlighted in bold. In what follows, ChatGPT response in Figure 3a will be called basic, and the responses in Figures 3b, 3c and 3d will be called comparable. When compared with the basic one, the response in Figure 3b contains explanatory items, the content of which includes 9 times a new parameter, namely the stroke. Although this parameter is included in the list of CVD risk factors, ChatGPT response does not explain its negative impact on the cardiovascular system. At the same time, as follows from the answers presented in Figures 3c and 3d, the parameter related to a person's age, under and over 60 years, was considered as a risk factor and ChatGPT generated a separate explanatory item for this parameter. Moreover, in the response presented in Figure 3d, clarifications regarding elderly people were included in the text of some explanatory items, since these explanations concern those people who are over 60 years old.

Based on 30 different ChatGPT prompts and analysis of the generated explanatory elements, a conclusion can be drawn. ChatGPT produces correct and consistent explanations of CVD risk factors. Although this LLM sometimes does not provide a separate explanation for a non-modifiable risk factor, it does use it to clarify the explanation for other CV risk factors in the parameter list. This type of explanation appears to be acceptable and adequate in the context of explanatory items *Expl* of proposed recommender algorithm.

## 7. Verification and Evaluation of the Recommendation Algorithm

The proposed CVD-Risk-Prevent algorithm refers to the knowledge-based recommendation algorithms, since it is based on explicit knowledge about person, about recommendation items and about their relations. Unlike other recommender algorithms, it does not depend on large datasets [20,21], and do not suffer from problem of "cold start". On the other hands the application of knowledge-based recommendation algorithms in domain of supporting adults in self-manage their CVD risk factors at home is new. Based on the above, our approach to evaluating the effectiveness of the proposed CVD-Risk-Prevent algorithm is based on verification methods commonly used in information and computer science. Therefore, this field of research suffers from a lack of datasets to study the performance of knowledge-based recommendation algorithms.

By verification of the algorithm, we mean (1) formal prof of algorithm correctness; (2) testing of algorithm; (4) pilot study of user's satisfaction in recommendation generated by algorithm; (5) assessment of the algorithm, recommendation items and texts of recommendations by doctors; (6) comparison with similar knowledge-based recommendation algorithms. This evaluation scenario is appropriate for knowledge-based recommendation algorithms and is often used as noted in [13].

*7.1. Formal Proof of the Correctness of CVD-Risk-Prevent Algorithm*

To proof of the correctness of the proposed algorithm Hoare logic with the Hoare triple was used. A triple describes how the execution of a piece of code (or algorithm) changes the state of the computation. A Hoare triple is of the form {{P}C{Q}}, where P and Q are assertions and C is a command [39]. A variable P is denoted the precondition and Q is the postcondition: when the precondition is met, executing the command establishes the postcondition. Assertions are formulae in predicate logic.

| $params$ $= high\ blood\ pressure, physical\ inactivity,$ $smoking$ | $params = high\ blood\ pressure, physical$ $inactivity, smoking, stroke$ |
|---|---|
| Controlling cardiovascular (CV) risk factors is crucial for maintaining heart health and reducing the risk of heart disease. Here are concise explanations for each of the specified risk factors:<br><br>### 1. ** High Blood Pressure**<br>1. High blood pressure can strain your heart and arteries, making them work harder and less efficiently.<br>2. Over time, the force of high blood pressure can damage the delicate tissues inside the arteries, leading to plaque buildup.<br>3. This buildup can narrow and block arteries, leading to a heart attack, stroke, or other serious heart conditions.<br>4. Controlling blood pressure can significantly lower the risk of heart failure and prolong overall life expectancy.<br><br>### 2. ** Physical Inactivity**<br>1. Regular physical activity helps strengthen the heart muscle, improves blood flow, and helps regulate weight.<br>2. Being inactive increases the risk of fatty material build-up in your arteries, which can lead to coronary artery disease.<br>3. Exercise helps lower blood pressure and reduce stress on the cardiovascular system.<br>4. Staying active can decrease the likelihood of developing other conditions that put strain on the heart, like diabetes and high cholesterol.<br><br>### 3. ** Smoking**<br>1. Smoking damages the lining of your arteries, leading to a build-up of fatty material that narrows the artery.<br>2. The chemicals in tobacco smoke increase the risk of blood clots forming and sticking to artery walls.<br>3. Smoking also causes plaque buildup in the arteries, increasing the risk of sudden blockages and heart attacks.<br>4. Quitting smoking can rapidly improve your heart health, reducing your risk of cardiovascular diseases significantly. | Controlling cardiovascular (CV) risk factors is **especially** crucial for someone with a **history of stroke** to prevent further heart-related issues. Here's why managing each risk factor is important:<br><br>### 1. ** High Blood Pressure**<br>1. High blood pressure can damage blood vessels, making them more susceptible to the formation of clots, which can lead to another **stroke**.<br>2. It puts additional strain on your heart, forcing it to work harder to pump blood, which can weaken the heart over time.<br>3. Controlling blood pressure reduces the risk of further damage to your cardiovascular system and can prevent complications like heart attack and kidney disease.<br>4. Effective management of blood pressure can significantly lower the chances of a **recurrent stroke** and other heart-related problems.<br><br>### 2. ** Physical Inactivity**<br>1. Regular exercise helps improve heart efficiency, lowers blood pressure, and enhances circulation, reducing the risk of **another stroke**.<br>2. Physical activity helps manage weight and reduces the likelihood of developing other cardiovascular problems.<br>3. Being active boosts overall energy levels and aids in the management of stress and depression, which are common **post-stroke issues**.<br>4. Incorporating routine physical activity into your lifestyle can prevent the progression of other conditions that might compound **stroke recovery**.<br><br>### 3. ** Smoking**<br>1. Smoking further damages cardiovascular health by thickening blood and increasing plaque buildup in arteries, leading to blockages.<br>2. It also increases the risk of **recurrent strokes** by impairing oxygen levels and blood flow throughout the body.<br>3. Quitting smoking can rapidly decrease the **risk of future strokes** and improve overall cardiovascular health.<br>4. Avoiding tobacco can also enhance the effectiveness of medications used to manage **other stroke-related health issues**. |
| a) | b) |

| $params = high\ blood\ pressure, age\ is\ under\ 60,$ $physical\ inactivity, smoking, stroke$ | $params = high\ blood\ pressure, age\ is\ over\ 60,$ $physical\ inactivity, smoking, stroke$ |
|---|---|
| It's vital for someone under 60 with a history of stroke and other cardiovascular risk factors to actively manage their health to prevent further complications. Here's a detailed explanation for each specific risk factor:<br><br>### 1. ** High Blood Pressure**<br>1. High blood pressure can cause further damage to your cardiovascular system, increasing the risk of a **second stroke or heart attack**.<br>2. It strains the heart by forcing it to work harder to pump blood, potentially leading to heart failure over time.<br>3. Managing blood pressure can prevent the progression of damage to your blood vessels and heart, reducing the likelihood of cardiovascular complications.<br>4. Keeping blood pressure in check can also improve overall longevity and quality of life, especially **after having a stroke**.<br><br>### 2. ** Age under 60**<br>1. **Being under 60** means you have the potential to significantly reduce long-term cardiovascular risks through lifestyle changes.<br>2. Early intervention in managing risk factors can lead to better health outcomes and a reduced risk of premature heart disease.<br>3. Younger individuals often recover faster and more effectively from lifestyle modifications than older adults.<br>4. Taking control of cardiovascular health now can preserve heart function and prevent disease progression as you age.<br><br>### 3. ** Physical Inactivity**<br>1. Lack of physical activity can lead to poor cardiovascular health, worsening outcomes for someone with a **history of stroke**.<br>2. Regular exercise helps lower blood pressure, improve cholesterol levels, and strengthen the heart muscle.<br>3. Staying active reduces the risk of obesity, diabetes, and further **strokes**—all common complications in inactive individuals.<br>4. Exercise also boosts mental health and mobility, which are crucial for recovery and quality of **life post-stroke**.<br><br>### 4. ** Smoking**<br>1. Smoking damages the blood vessels and increases the risk of clot formation, which can lead to **another stroke** or heart attack.<br>2. It diminishes overall cardiovascular health and reduces the effectiveness of **stroke recovery treatments**.<br>3. Quitting smoking can dramatically reduce the likelihood of **recurrent strokes** and other cardiovascular diseases.<br>4. Stopping smoking improves lung function and oxygenation, vital for overall health and especially crucial **after a stroke**. | For a person over 60 with a history of stroke and other cardiovascular risk factors, managing these factors is critical to maintaining health and preventing further complications. Here are explanations tailored to each risk factor:<br><br>### 1. ** High Blood Pressure**<br>1. High blood pressure can accelerate the wear and tear on blood vessels, increasing the risk of a **second stroke** and other heart-related issues.<br>2. It forces your heart to work harder, which can lead to heart enlargement and eventual heart failure.<br>3. Keeping blood pressure controlled helps to prevent further damage to your cardiovascular system, reducing the **risk of heart attack** and kidney disease.<br>4. Effective management of blood pressure is crucial for maintaining quality of life and reducing the risk of mortality from heart conditions.<br><br>### 2. ** Age under 60**<br>1. As you age, your cardiovascular system naturally faces more stress, making it more important to control risk factors.<br>2. Older adults are more susceptible to the effects of high blood pressure and other stressors on heart health.<br>3. Proactive management of cardiovascular health can lead to improved longevity and a better quality of life in later years.<br>4. Early and consistent intervention can prevent rapid deterioration of health, helping to maintain independence and reduce the burden of disease.<br><br>### 3. ** Physical Inactivity**<br>1. Inactivity can lead to the weakening of the heart muscle and poorer blood flow, increasing the risk of heart disease and **a second stroke**.<br>2. Regular physical activity helps manage weight, reduces hypertension, and improves lipid profiles, which are crucial for someone with a **stroke history**.<br>3. Exercise enhances mobility and balance, reducing the risk of falls and other physical complications that can be devastating **for older adults**.<br>4. Staying active is also beneficial for mental health, helping to combat **post-stroke depression** and cognitive decline.<br><br>### 4. ** Smoking**<br>1. Smoking continues to damage blood vessels and increases the risk of forming clots, which could lead to **additional strokes** or heart attacks.<br>2. It impairs lung function, reducing oxygen available for vital organs and issues, which is especially detrimental for seniors.<br>3. Quitting smoking can lead to immediate improvements in heart rate and blood pressure, significantly benefiting overall cardiovascular health.<br>4. Stopping smoking reduces the progression of cardiovascular deterioration, which is **critical for aging individuals** to maintain their health. |
| c) | d) |

Figure 3. Examples of ChatGPT-4 responses on prompts with different parameters

Let us determine the initial conditions for the input data contained in the sets X, Z, R. The set of indicators of CV health X is defined in expression (12) and is presented in Table 2. Note that $\forall x_j \geq 0$, $j = 1,2 \ldots 17$, while $x_j = 0$ means that the user did not enter data or entered data corresponding to the absence of CV risk factors. The model of CV risk factors Z is defined in expression (7). Content and designation of CV risk factors are presented in Table 1, which provides threshold p values for some risk factors. The set R contains a set of recommendation items of different types, used to generation a person recommendation.

Let us denote P as the precondition for the CVD-Risk-Prevent algorithm that corresponds to the truth of the conditions on the input data given above. We assume that the assertion P in which each free occurrence of input variables has been replaced by the values of corresponding to sets X, Z, R. The result of the CVD-Risk-Prevent algorithm is the text of the recommendation R*, which combines recommendation items of various types presented in formulas (28-31), while $ESG \in R$ depends on the person's class and his predictive assessment of total CV risk C, the components $EsR \in R$, $Inf \in R$, $Expl \in R$, $Plan \in R$ are associated with the person's risk factors.

Let us denote D as logical expression $\{ESG \neq \emptyset\}$, which is obtained as a result of applying Rule 7 (see expression 32), the truth of this expression relies that the $ESG \in R$ is not empty and is contained in the text of recommendation R*. Then we denote logical expression $W(i) = \{EsR(i) \cup Inf(i) \cup Exp;(i) \cup p(i)\}$, which is derived according to Rule 8 for each person's existing CV risk factor (see expression 33).

Introduce the variable W which expresses the logical statement that the recommendation items $EsR \in R$, $Inf \in R$, $Expl \in R$, $Plan \in R$ for each user CV risk factor are exist in the text of recommendation R*:

$$W = \bigvee_{i=1}^{13} W(i). \qquad (34)$$

Let us define the postcondition of the algorithm in the following form of logical statement which will be truth then and only then if expressions D and W are truth, which means the algorithm produces the recommendation R*

$$V = \{D \vee \{W \neq \emptyset\}\}. \qquad (35)$$

Theorem 1. Given precondition P is true the statement (35) is true for CVD-Risk-Prevent algorithm. It means we need to prove that {P} S {V} is true, in other words if P, then V is true. Here S denotes the rules used in the proposed algorithm.

The proof of the formulated Theorem 1 is given in Appendix A.

## 7.2. Testing of CVD-Risk-Prevent Algorithm

To test the proposed recommendation algorithm, an experimental software system in the form of web-service was designed to support adults in self-manage of their CVD risk factors at home. We tested the developed CVD-Risk-Prevent algorithm on previously defined manual example with data represented in Table 3 since this algorithm is deterministic in its nature.

Step-by-step execution of the algorithm. At the first stage, the recommendation algorithm analyzes the person's profile and, using the CV risk factor model Z, generates the vector $Factor = (0,0,0,0,0,0,0,0,1,0,1,1,1)$. At the second stage, using rules that analyze the values of the components of the $Factor$ vector, the person's $Class = (0,1,1,0,0)$ is created, since Rule2 and Rule3 are triggered. At the third stage, the total CVD risk is assessed as $Sco = f(1) \vee f(2) \vee f(4) \vee f(5) = 0$ using SCORE2/SCORE2-OP model given in [4, 6].

Table 3. Example 1 of CV health indicators of a person No.1

| X | Indicator | X | Indicator |
|---|---|---|---|
| X1 | Gender – female | X10 | Cholesterol(total) – 5.0 mmol/l |
| X2 | Age – 49 years | X11 | Non-high density lipoprotein cholesterol – 3.0 mmol/l |
| X3 | Height – 170 cm | X12 | Systolic blood pressure – 160/90 mmHg |
| X4 | Weight – 74 kg | X13 | Glucose level – 4.8 mmol/l |
| X5 | Family history of CVD – No | X14 | Physical inactivity –Yes |
| X6 | Presence of CVD - No | X15 | Smoking – Yes |
| X7 | Chronic kidney disease – No | X16 | Unhealthy diet – Yes |
| X8 | History of CV events – No | X17 | Symptoms of angina pectoris, with a significant deterioration in health – No |
| X9 | Diabetes mellitus type 2 – No | | |

As a result, a predictive estimate of the total CV risk = 6% was obtained, and the linguistic meaning of the level of CV risk for person No. 1 was assessed as C = "high". At the fourth stage, recommendation items are generated that determine the content of multidimensional recommendation. Since this person belongs to classes 2 and 3, the person has modifiable biological and behavioral risk factors, which is reflected in the text of person's recommendation presented below:

" Goal: CVD prevention is your primary goal. Considering your risk factors, you are advised to consult a therapist as planned to receive medical care and possibly additional examination. It is necessary to lower blood pressure, reduce salt intake and stop smoking. Target values for your age: blood pressure no more than 130 mmHg, the total amount of salt consumed in any form is no more than 5 grams per day (1 teaspoon). Set a goal to quit smoking in 10 days.

Information: Your total CV risk in a ten-year perspective is classified as high. Your cardiovascular health can be improved by controlling your blood pressure, diet and quitting smoking.

Explanation: Please note that elevated blood pressure (BP) may be a sign of incipient hypertension. Systematic elevation of blood pressure is a major risk factor for heart disease, stroke and other serious health problems. As blood pressure rises, the load on the heart and blood vessels increases, and the heart begins to work harder. The more salt a person consumes, the higher the risk of premature death from any cause. For the average 50-year-old woman, excess salt steals 1.5 years of life and significantly increases the risk of arterial hypertension.

2. Smoking is one of the main factors of high risk of death from CVD and other chronic diseases. Quitting smoking can reduce this CVD risk by half. Those who do not smoke have an average life expectancy of 10 years longer. The risk of CVD in smokers under 50 years of age is five times higher than in non-smokers. After just a few puffs, the heart begins to wear out - carbon monoxide from tobacco smoke sharply reduces the concentration of oxygen in the blood, and it has to be distilled in large volumes to prevent the body from feeling a lack of oxygen.

Plan of actions: Did you know that 90% of people who promise themselves to become better do not keep this promise because most people do not understand the process of creating sustainable results, so they never take the first step. An effective action plan will give you the start and the driving force that will lead you to improve the quality and duration of your active life:

A. Consultation with a therapist. To control your blood pressure, preventative medications are necessary, as determined by your doctor.

B. Imagine what would happen if you didn't smoke at all. Would you be reaching for a cigarette right now? Probably not. That's true for the vast majority of people: it's rare to start smoking after the age of 25. What would your life be like if you didn't smoke? Would you think about cigarettes all the time or would you be more free?

Step 1: Make a decision to quit smoking and stick to it.

Step 2: Set a date in advance when you will quit smoking. It can be timed to coincide with a significant event (vacation, birthday, vacation). In the future, don't look for excuses to postpone the planned date.

Step 3. Visit theaters, cinemas, and museums more often, in the halls of which smoking is prohibited.

Step 4: Bet someone that you will achieve your goal.

Step 5. Quit smoking for one day, tomorrow for another, the day after tomorrow for another, etc. Be happy with every unsmoked cigarette.

Step 6: Find a partner to quit smoking with. Try to help someone else quit smoking."

As follows from the recommendation, generated by CVD-Risk-Prevent algorithm, its text covers all person CVD risk factors, contains recommendation items corresponding to multidimensional model of recommendations R (1) and satisfies the requirements for recommendations given in Section 3.

*7.3. Study of user satisfaction in recommendations received*

To understand the effectiveness of the proposed CVD-Risk-Prevent algorithm from the user's perspective, additional study was conducted involving 15 independent persons who provided consent for the use of CV health indicators. The age of the participants varied from 34 to 70 years, among them 53% were men, 47% were women. The purpose of the study is to find out the degree of user satisfaction with the content of the recommendations received and the degree of agreement of their opinions. To obtain recommendations in this study a software implementation of the CVD-Risk-Prevent algorithm was developed, a fragment of the user interface for input CV health indicators is depicted in Figure 4. To analyze user opinions, a special questionnaire was developed, the questions of which were defined in such a way as to evaluate the features of the received recommendation for compliance with requirements represented in Section 3. Likert scale was used to create a survey of persons' satisfaction with the content of the recommendations received. Users assessed the individual recommendations according to 7 criteria, using score from 1 to 5, where score 1 corresponds to completely disagree and 5 – to completely agree with the statement given in the questionnaire.

Figure 4. Fragment of the user interface for entering CV health indicators

The results of the user survey based on developed questionnaire can be seen in Table 4. A degree of user satisfaction (DUS) is calculated by dividing the mean by the maximum score, i.e. 5:

$$DUS_l = \frac{1}{5k}\sum_{v=1}^{k} y_{lv},$$

here $y_{lv}$ denotes the score of the v-th participant on the l-th question of the questionnaire, and k is the number of participants.

Table 4. Questionnaire for user satisfaction. Here SD designates standard deviations DUS denotes degree of users' satisfaction

| No. | Feature of the individual recommendations | Mean | SD | DUS |
|---|---|---|---|---|
| 1 | Completeness. The resulting recommendations comprehensively cover your cardiovascular health concerns and inform you about cardiovascular risk factors | 4.5 | 0.28 | 0.9 |
| 2 | Explainability. Recommendations explain the impact of risk factors on cardiovascular health | 4.5 | 0.28 | 0.9 |
| 3 | Timeliness. The recommendations received are relevant for seeking medical help. | 4.4 | 0.49 | 0.88 |
| 4 | Accuracy and individual focus. The recommendations received are specific to you and presented in understandable language. | 4.2 | 0.4 | 0.84 |
| 5 | Planning. A recommended plan of steps you need to take to help you achieve your cardiovascular disease prevention goals. | 4.6 | 0.49 | 0.92 |
| 6 | Motivation. The recommendations you receive will motivate you to manage your CV health. | 4.2 | 0.4 | 0.84 |
| 7 | Utility. The recommendations received are useful for you | 4.8 | 0.18 | 0.96 |
|  | Mean | 4.46 | 0.36 | 0.89 |
| Cronbach's alpha = 0.79 | | | | |

To determine the consistency of the survey results, Cronbach's alpha statistics were calculated using MS Excel. A Cronbach's alpha value of 0.79 indicates acceptable user agreement on the quality of recommendations received. Data in Table 4 show that participants agree that individual recommendations meet all quality criteria. Moreover, some persons noted that they had previously received exactly such recommendations from their attending physicians. An analysis of the average ratings of the features of recommendations shows that the Utility property received the highest rating, while properties such as Planning and Motivation received the lowest rating. Interestingly, the explainability of individual recommendations, for the generation of which Chat was used, received a high score of DUS = 0.9.

The difference in the degree of satisfaction with recommendations for self-management of CVD risk in male and female is shown in Figure 5.

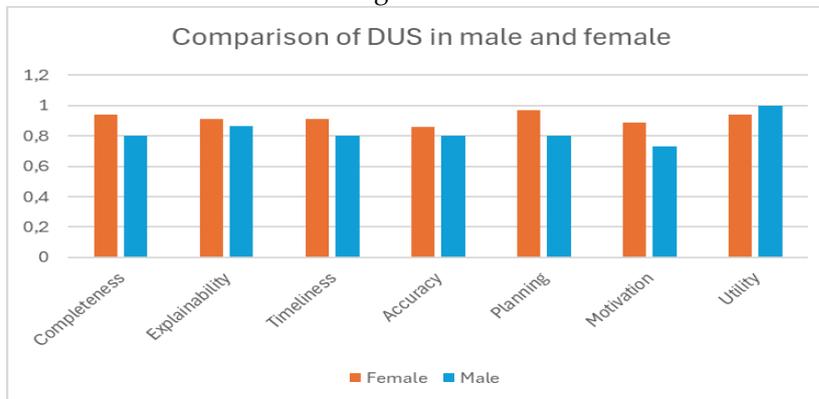

Figure 5. Comparison on degree of satisfaction among male and female recommendation received

As follows from the Figure 5, men on average rated the feature of recommendations slightly lower on almost all features than women. At the same time, men rated the utility of recommendation higher than women. Considering the results obtained using the CVD-Risk-Prevent algorithm and Cronbach's alpha coefficient of 0.79, it can be concluded that persons perceive the developed recommendation algorithm as a possible tool for supporting adults in self-management of CVD risk factors.

*7.4. Expert Evaluation*

Two independent experts specializing in the prevention and treatment of CVD gave their consent for a qualitative assessment of the proposed CVD-Risk-Prevent algorithm from the point of view of evidence-based medicine.

The first expert is a Doctor of Medical Sciences, professor of the medical faculty of the Institute of Medicine, Ecology and Physical Culture of Ulyanovsk State University, Russia, the second one is a cardiologist and therapist working at the Moscow clinic GLOBAL MEDICAL SYSTEMS (GMS LLC), Russia. Both specialists have extensive experience, more than 20 years of practice in the treatment of persons with CVD, as well as their preventive counseling. The description of the algorithm, the models, the rules it uses, information sources, person's profiles as well as the resulting texts of recommendations were independently analyzed by experts. The experts focused on assessing the correctness of recommendations obtained by the CVD-Risk-Prevent algorithm.  the correctness of recommendation was considered as their compliance with the person's profiles, their individual CV risk factors, and official guidelines [6, 7].

Both experts expressed the opinion that the proposed algorithm creates correct recommendations that can be used by adults in self-monitoring CVD risk factors and gave a positive assessment of the content and structure of multidimensional recommendations based on usefulness and safety criteria.

Doctors agree that the proposed algorithm is also promising in clinical practice, since it can reduce the doctor's time when preparing recommendations for the patient. At the same time, experts made some useful comments. In particular, to support adults with cardiovascular disease at home, the content of recommendations should be more specific and could be expanded using electronic health record data. For example, it would be useful for such patients to add recommendations for taking prescribed medications or a description of the dynamics of CVD health indicators.

*7.5. Comparison with Similar Knowledge-based Recommendation Algorithms*

According to [40] proposed CVD-Risk-Prevent algorithm has several key advantages as knowledge -based recommendation algorithm. The first advantage is that it overcomes the data shortage obstacles that plague collaborative and content filtering recommendation algorithms. Compared to collaborative algorithms and content filtering, the proposed algorithm does not require many users, does not require any datasets [20]. It's applicable to one person because it uses knowledge in the form of rules similar to those that doctors use in their practice using modern clinical guidelines. Another benefit of CVD-Risk-Prevent algorithm is the ability to leverage the integral expertise provided in guidelines and clinical practice recommendations using cause-and-effect statements. The novelty of the proposed algorithm from knowledge-based recommendation algorithms lies in its focus on support adults in self-management of their CVD risk factors at home based on official guidelines, in the multidimensional recommendation model, in the original model of user profile including predictive assessments of CVD risk, and in the method of generating recommendations combining rules and LLM, in particular, ChatGPT. To highlight the opportunity of CVD-Risk-Prevent algorithm the comparison with similar knowledge-based recommendation algorithms was done.

Table 5. The comparison health knowledge-based recommendation algorithms

| Criteria of comparison | | Spoladore, D. 2021 | Lopez-Barreiro, J, 2023 | Wang Z., 2020 | Proposed |
|---|---|---|---|---|---|
| Perspective of comparison | Indicator of comparison | | | | |
| Number of risk factors in User profile corresponding to CVD risk factors | Lifestyle | 2 | 9 | 4 | 3 |
| | Biological | - | - | 4 | 5 |
| | Associated with non-modifiable factors | - | - | 4 | 5 |
| | Descriptive estimates | 2 | 9 | 12 | 13 |
| | Predictive estimates | - | - | - | 1 |
| Recommendation associated with self-management of CVD risk factors | User focus | One class. Older adults with chronic pathologies | One class. Persons without medical conditions | One class. Chronic disease patients, including CVD | Five classes including adults with or without any CVD risk factors as well symptoms of angina |
| | Goal | Support healthy diet and physical activity | Support physical, mental, and nutritional health | Support selection of educational materials | Support self-management of CVD risk factors |
| | Results | 32 dishes, diet plan for each day | Ranked list of challenges, selected from 30 health challenges | Ranked list of educational materials | Goal, information, explanation and plan for each person's risk factor |
| | Knowledge models and algorithms | Ontology, expert knowledge | Expert ranking model | Rules, ontology, natural language processing | Rules, ChatGPT |
| | Guidelines or other official materials used in clinic practice | ACSM guideline, clinical literature | SF-36, sHEI-15 and OSC questionnaires | Corpus of educational materials | 2021 ESC Guidelines on CVD prevention in clinical practice, SCORE algorithms |
| | Evaluation | Two use cases | Four experts, 30 uses | MAE on key words for corpus of educational materials, | Formal prof, testing on step-by-step example, two experts, 10 users |

| | | | | marked by two experts | |

As follows from Table 5, the proposed algorithm estimates a larger number of CVD risk factors in user profile to construct recommendations, although it is inferior to other algorithms in terms of lifestyle factors. At the same time, only the proposed algorithm takes into account a predictive assessment of the total CVD risk, while in the compared algorithms the user profile contains only descriptive assessments. It follows from this that in the proposed algorithm, the user profile can describe more different states of a person's CV health, namely, in the amount of 2^14. As a consequence, this provides a greater diversity of recommendations generated by the proposed algorithm than in the compared ones. Comparative perspective on recommendation for self-management of CVD risk factors includes user focus and content of recommendation. In the articles [9,23,28] the recommendations are intended for one class of users, namely, adults with chronic pathologies or persons without medical conditions, which is narrower than in the proposed algorithm, since in addition to such users, recommendations are intended for five different classes, including adults with or without any CVD risk factors as well symptoms of angina. Regarding the content of the recommendations, in contrast to one-dimensional recommendations presented as a list of dishes, exercises [28], challenges [23] or educational materials [9], the proposed recommendations are multidimensional, the content of which is revealed in four aspects for each person's risk factor: targeted, informational, exclamational and planning. Information and semantic capacity are what distinguishes the content of the proposed multidimensional model of recommendation from those compared.

As shown in Table 5, the goal of generating recommendations to support adults in self-management of CVD risk factors at home based on official guidelines [6,7] was formulated only in the proposed algorithm. Moreover, to the best of our knowledge, this goal and its corresponding results have not been reported in research on recommendation algorithms. In our opinion, the reason is that the field of disease prevention, including CVD, is a new area for recommendation algorithms with its own challenges and opportunities. However, it should be noted that official materials widely used in clinical practice were used to construct ontologies [28] or to obtain information on lifestyle indicators [23].

The main difference in the compared knowledge-based recommendation algorithms is in the models used. Ontologies as a knowledge representation model are a suitable model in knowledge-based recommendation algorithms, since they allow us to formalize the relations of concepts at different levels. At the same time, from the point of view of transparency, the complexity of creation and explainability of ontologies with SWRL rules are inferior to "If-Then" rules. The proposed algorithm uses rules of 3 types: rules for identification of person CV risk factors, rules to assess person class and predictive score of total CV risk level, and rules to generate the recommendations. In the CVD-Risk-Prevent algorithm, for the first time, the rules for generating recommendations content are combined with a large language model, namely ChatGPT, for greater diversity in explanatory items.

The conducted satisfaction analysis both from the point of view of end users and from the point of view of doctors showed that the explanatory items in the recommendations created using ChatGPT-4 received good marks expressed in degree of user satisfaction to the formulated requirements for the recommendations. The formal proof of the correctness of the proposed CVD-Risk-Prevent algorithm is its other important difference in the evaluation process from the recommendation algorithms being compared.

## 8. Discussion and Conclusion

In the article knowledge-based recommendation CVD-Risk-Prevent algorithm for supporting self-management adults at home was proposed and studied. This recommendation algorithm combines LLM and rules to make multidimensional recommendations by converting general clinical practice, CVD prevention guidelines [6,7] and open world knowledge into rule-based logic. To develop CVD-Risk-Prevent algorithm 8 requirements to recommendation were defined, which determine the structure of proposed formal model of recommendation content including explanations. The latter is crucial for HRS, in particular, for CVD prevention in adults at home where explaining the recommendations can help make the right decision. The capabilities of ChatGPT have been examined and successfully used to create explanatory items. The authors are not aware of methods, algorithms or HRSs for generating recommendations, with such features. CVD-Risk-Prevent algorithm can be implemented as a separate service in the form of a mobile [42] or web application [43], interacting with both the user and smart digital medical gadgets. As noted in the WHO Global Strategy and Guidelines for Digital Health, the use of digital and information technologies, including artificial intelligence, data analytics, and software tools, smartphone applications and health websites, has proven its potential to improve health, health literacy and disease prevention for each person, as well as among the population as a whole [44]. The main benefits of the proposed recommendation algorithm for adult population are as follows:

a. Involving the population in taking care of their health, improving the quality and duration of active life.
b. Increasing medical literacy, allowing you to make optimal decisions in the direction of managing CV risks.
c. Ensuring monitoring and control of CV risk factors.
d. Timely seeking medical help if symptoms of CVD or sudden deterioration occur.
e. Reducing financial and time costs associated with visiting hospitals and providing high-tech medical care in the treatment of CVD.
f. Increasing the availability of preventive care for the population 24/7 and increasing the percentage of the population covered by prevention of CVD based on information technologies and AI.

The CVD-Risk-Prevent algorithm proposed in this article does not require training to calculate the optimal parameters of models from many examples, as it is based on proven medical knowledge and pre-training large language model, namely ChatGPT. Verification and evaluation showed that the CVD-Risk-Prevent algorithm outperforms similar knowledge-based recommendation algorithms in terms of the diversity of CVD risk factors contained in the user profile and in terms of recommendation content having four dimensions of meaning. The proposed recommender algorithm received a positive qualitative assessment from experienced therapists and suggestions for its further development. End users assessed the recommendations generated by CVD-Risk-Prevent algorithm for compliance with the formulated requirements, which made it possible to calculate the degree of their satisfaction with the recommendations and the degree of consistency of their opinions. The results showed good agreement among end users and a high degree of satisfaction with the recommendations, including explainable items obtained using MML with a score of 0.9. Let us note the challenges in developing and verification of CVD-Risk-Prevent algorithm: (1) the lack of open datasets of recommendations for self-management of CVD risk factors, and challenges in collecting implicit user data [18]; (2) when using open datasets for validation and study some features of new algorithm, research is limited to the set of health indicators that are given in these datasets. The contributions of the research we defined as follows:

1. To overcome limitations and fill gaps in research on knowledge-based HRS, two formal models was presented: the CVD risk factor model and the multidimensional recommendation model. The risk factor model includes a full set of CVD risk factors given in the 2021 clinical guidelines. The structure of multidimensional recommendations combines 4 types of recommendation items similar to those provided to the person during preventive consultation: targeted, informational, explanatory and planned. These models, being the basis of the proposed recommendation algorithm, are used to analyze the CV status of the person's health, and to synthesize his individual recommendations, for which three types of rules were developed: rules for identifying risk factors for cardiovascular diseases in a person, rules for assessing the person's class, and its overall level of CVD risk, and the rules used to create structured recommendations for CVD prevention.

2. The novelty of the proposed algorithm from knowledge-based recommendation algorithms lies in its focus on supporting self-management of CV risk factors in adults at home, in the original recommendation model, in the completeness of the set of CVD risk factors, and in the method of generating recommendations, which combines rules and LLM.

A limitation of the proposed recommendation algorithm is the lack of recommendations for the use of medications, such as statins or aspirin. It is assumed that the appointment of such an intervention is the task of the attending physician. The lack of analysis of tendencies in the person's profile and his CV risk factors is another limitation of the proposed recommendation algorithm. In our opinion, a common limitation of knowledge-based recommendation algorithms, including the one proposed in this article, is the use of subjective quality assessments. CVD-Risk-Prevent algorithm does not consider such risk modifiers as psychological stress, ethnicity, imaging data and comorbidity. Another limitation relies at focusing on European adult population and on prevention only CVD. Although the proposed recommendation algorithm is deterministic, we recognize that larger studies involving medical professionals and different types of users are needed to evaluate its effectiveness in preventing CVD.

The recommendation system based on the proposed algorithm, as can be concluded from the results of the survey of future users conducted in the article, will be in demand both among the adult population for managing cardiovascular health at home and among doctors. Therefore, further research will be focused to the developing recommendation system for supporting adults in self-management of CVD risk factors at home and its experimental studies to obtain numerical performance estimates for CVD-Risk-Prevent algorithm. To expand the opportunities of the software implementation of the proposed recommendation algorithm, research will be conducted in four directions: 1) developing an interface for integration with digital gadgets, medical information systems and electronic medical records; 2) exploring ways to integrate the proposed algorithm with recommendation algorithms for collaborative and content filtering when a large database of persons is created; 3) solving problems related to ethical issues and confidentiality; 4) the use of natural language processing and artificial intelligence technologies to automate the addition of knowledge about new risk factors from new guidelines.

Author Contributions: methodology, writing, visualization, T.V.A; formal analysis, data curation, P.V.P.; writing, original draft preparation and editing A.I.M. All authors contributed equally. All authors have read and agreed to the published version of the manuscript.

Acknowledgements. The authors express their gratitude to the experts for their careful analysis and valuable comments, which undoubtedly improved the content of this article: Doctor of Medical Sciences, Professor, Head of the Department of Therapy and Occupational Diseases, Faculty of Medicine, Institute of Medicine, Ecology and Physical Culture, Ulyanovsk State University, Russia, and Yuri Sergeyevich Zamashkin, cardiologist in GLOBAL MEDICAL SYSTEMS Limited Liability Company (LLC GMS), Moscow, Russia.

Funding: This research was performed in the framework of the state task in the field of scientific activity of the Ministry of Science and Higher Education of the Russian Federation, project "Models, methods, and algorithms of artificial intelligence in the problems of economics for the analysis and style transfer of multidimensional datasets, time series forecasting, and recommendation systems design", grant no. FSSW-2023-0004.

Institutional Review Board Statement: Not applicable.
Informed Consent Statement: Not applicable.
Data Availability Statement: Not applicable.
Conflicts of Interest: The authors declare no conflict of interest.

**Appendix A.**

Appendix A provides a formal proof of Theorem 1 formulated in Section 7.1. We will carry out the proof for each stage of the CVD-Risk-Prevent algorithm given in Section 5.

STEP 1. The result of the first Stage of the algorithm is the Factor vector (14).

Let us denote the postcondition U for the output data in the form of a statement that each component of the Factor vector is $f(i) \in \{0,1\}, i = 1,2 \dots 13$.

Let us prove {P} S {U}, S is the set of transformations of this stage of the algorithm.

According to expressions (15) and (16), taking in account that each of x(5), x(6), x(7), x(8), x(9), x(14), x(15), x(16), x(17) ∈ {0,1}:

$\{P\}\ S1\ \{U1\},\ U1 = \{f(k) \in \{0,1\}, k = 1,2,3,4,5,11,12,13\}$.

Following expression (17):

$\{U1\}\ S2\ \{U2\}, U2 = \{f(k) \in \{0,1\}, k = 1,2,3,4,5,6,11,12,13\}$.

Using expression (18):

$\{U2\}\ S3\ \{U\}, U = \{f(k) \in \{0,1\}, k = 1,2,\dots,13\}$.

Therefore, as $\{P\}\ S1\ \{U1\}, \{U1\}\ S2\ \{U2\}, \{U2\}\ S3\ \{U\}$ then {P} S {U} is true.

STEP 2. Let us denote by U the preconditions of second Stage of algorithm. The result of the second Stage of the algorithm is the vector $Class = \{Class(1), Class(2), Class(3), Class(4), Class(5)\}$. Let us denote the postconditions Q on the output data in the form that each Class(i)∈{0,1} and $\exists i \in \{1,2,\dots,5\}$, such that Class(i)=1. Let us prove {U} S {Q}, S is the set of transformations of this stage of the algorithm. To prove this, apply the rules Rule 1,…,Rule 5 given in expressions (20-24) to the Factor vector. Since $f(i) \in \{0,1\}, i =$

1,2,...,13, for convenience, we replaced the maximum function in these rules with logical multiplication, since this replacement does not violate the result of the rules Rule 1,…,Rule 5.

Then in respect to expressions (20-24):

$\{U\} \wedge \{\wedge_{i=1}^{13} f(i) = 0\}\ S4\ \{Q1\},\ Q1 = \{Class(1) = 1\}$,

$\{U\} \wedge \{\wedge_{i=1}^{13} f(i) = 1\}\ S4\ \{\neg Q1\},\ \neg Q1 = \{Class(1) = 0\}$,

$\{U\} \wedge \{\neg Q1\} \wedge \{\wedge_{i=11}^{13} f(i) = 1\}\ S5\ \{Q2\},\quad Q2 = \{Class(2) = 1\}$,

$\{U\} \wedge \{\neg Q1\} \wedge \{\wedge_{i=11}^{13} fi = 0\}\ S5\ \{\neg Q2\},\ \neg Q2 = \{Class(2) = 0\}$,

$\{U\} \wedge \{\neg Q1\} \wedge \{\neg Q2 \vee Q2\} \wedge \{\wedge_{i=6}^{10} fi = 1\}\ S6\ \{Q3\},\ Q3 = \{Class(3) = 1\}$,

$\{U\} \wedge \{\neg Q1\} \wedge \{\neg Q2 \vee Q2\} \wedge \left\{\bigwedge_{i=6}^{10} fi = 0\right\}\ S6\ \{\neg Q3\}\ \neg Q3 = \{Class(3) = 0\}$,

$\{U\} \wedge \{\neg Q1\} \wedge \{\neg Q2 \vee Q2\} \wedge \{\neg Q3 \vee Q3\} \wedge \{\wedge_{i=2}^{5} fi = 1\}\ S7\ \{Q4\},\ Q4 = \{Class(4) = 1\}$,

$\{U\} \wedge \{\neg Q1\} \wedge \{\neg Q2 \vee Q2\} \wedge \{\neg Q3 \vee Q3\} \wedge \left\{\bigwedge_{i=2}^{5} fi = 0\right\}\ S7\ \{\neg Q4\},\quad \neg Q4 = \{Class(4) = 0\}$,

$\{U\} \wedge \{\neg Q1\} \wedge \{\neg Q2 \vee Q2\} \wedge \{\neg Q3 \vee Q3\} \wedge \{\neg Q3 \vee Q3\} \wedge \{f1 = 1\} S8\ \{Q5\},\quad Q5 = \{Class(5) = 1\}$,

$\{U\} \wedge \{\neg Q1\} \wedge \{\neg Q2 \vee Q2\} \wedge \{\neg Q3 \vee Q3\} \wedge \{\neg Q3 \vee Q3\} \wedge \{f1 = 0\} S8\ \{\neg Q5\},\quad \neg Q5 = \{Class(5) = 0\}$,

$\{\{Q1\} \wedge \{\neg Q2\} \wedge \{\neg Q3\} \wedge \{\neg Q4\} \wedge \{\neg Q5\}\} \vee \{\{\neg Q1\} \vee \{Q2\} \vee \{Q3\} \vee \{Q4\} \vee \{Q5\}\}\ S9\ \{Q\}$.

Therefore, {U} S {Q}. Since {P} S {U} and {U} S {Q}, then {P} S {Q}.

STEP 3. The precondition for Stage 3 of the algorithm is U. The result of the 3rd stage of the algorithm is the statement that there is a value C such that $C \in \{low, moderate, high, very\ high\}$. Let us denote the postconditions B for the output data in the form of a statement $B = B23 \vee B24 \vee B25 \vee B26$. Let's introduce logical statements

$$B23 = \{C = low\};\ B24 = \{C = moderate\};\ B25 = \{C = high\};\ B26 = \{C = very\ high\}.$$

Let us prove {U} S {B}, S is the set of transformations of this stage of the algorithm.

In respect to expressions (25-26)

$$B1 = \{Sco = f(1) \vee f(2) \vee f(4) \vee f(5)\};$$
$$B2 = \{CVrisk = FUNC(x(1), x(2), x(11), x(12), x(15))\};$$
$$B3 = \{CVrisk < 2.5\};\ B4 = \{2.5 \leq CVrisk < 5\};$$
$$B5 = \{5 \leq CVrisk < 10\};\ B6 = \{CVrisk \geq 10\}.$$

Then

$$\{U\} \wedge \{\neg B1\}\ S10\ \{B2\};\ \{B2\} \wedge \{B3\}\ S11\ \{B23\};$$
$$\{B2\} \wedge \{B4\}\ S12\ \{B24\};\ \{B2\} \wedge \{B5\}\ S13\ \{B25\};$$
$$\{B2\} \wedge \{B6\}\ S14\ \{B26\};\ \{U\} \wedge \{B1\}\ S15\ \{B26\};$$

$\{B23 \vee B24 \vee B25 \vee B26\}\ S16\ \{B\}$.

As {P} S {U} and {U} S {B} then {P} S {B}.

STEP 4. Let us denote the postconditions D for the output data in the form of a statement according to rule 7 from expression (32)

$$D = D1 \vee D2 \vee D3 \vee D4 \vee D5,$$
$$D1 = \{EsG(1) = true\};\ D2 = \{EsG(2) = true\};\ D3 = \{EsG(3) = true\};$$
$$D4 = \{EsG(4) = true\};\ D5 = \{EsG(5) = true\}.$$

Let us prove {Q} S {D}, S is the set of transformations of this stage of the algorithm.

$\{Q\} \wedge Class(1)\}\ S17\ \{D1\};\ \{Q\} \wedge \neg Class(1)\}\ S17\ \{\neg D1\}$;

$\{Q\} \wedge Class(2)\}\ S18\ \{D2\};\ \{Q\} \wedge \neg Class(2)\}\ S18\ \{\neg D2\}$;

$\{Q\} \wedge Class(3)\}\ S19\ \{D3\};\ \{Q\} \wedge \neg Class(3)\}\ S19\ \{\neg D3\}$;

$\{Q\} \wedge Class(4)\}\ S20\ \{D4\};\ Q\} \wedge \neg Class(4)\}\ S20\ \{\neg D4\}$;

$\{Q\} \wedge Class(5)\}\ S21\ \{D5\};\ \{Q\} \wedge \neg Class(5)\}\ S21\ \{\neg D5\}$;

$\{Q\} \wedge \{D1 \vee D2 \vee D3 \vee D4 \vee D5\} S22\ \{D\}$

Since, according to the truth of the expression Q, there exists $i$ such that $Class(i) = 1, i = 1,2,...,5$, then the component of recommendation $ESG \neq \emptyset$.

Consequently, {Q} S {D}, as {P} S {U} and {U} S {Q} and {Q} S {D}, then {P} S {D}.

Let us prof $\{U\}S\{W\}, \{W \neq \emptyset\}$, where $W$ defined using Rule 8 (see (33)) and the expression (34):

$\{U\} \wedge \{B\} \wedge \{f(1) = 1\}\ S23\ \{W1\},\quad W1 = \{EsR(1) \cup (C \cup Inf(1)) \cup Expl(1) \cup p(1)\}$,

$\quad\{U\} \wedge \{B\} \wedge \{f(2) = 1\}\ S24\ \{W2\},\quad W2 = \{EsR(2) \cup (C \cup Inf(2)) \cup Expl(2) \cup p(2)\}$

……………………………………………………………………………….,

$$\{U\} \wedge \{B\} \wedge \{f(13) = 1\} \, S35 \, \{W13\},$$
$$W13 = \{EsR(13) \cup (C \cup Inf(13)) \cup Expl(13) \cup p(13)\},$$
$$\{U\} \wedge \{B\} \wedge \{W1 \vee W2 \vee \ldots \vee W13\} \, S36 \, \{W\}.$$

Following the last statement, it can be concluded that $\{W \neq \emptyset\}$, if person has any CV risk factor.
Consequently, $\{U\} \wedge \{B\} \, S \, \{\{W \neq \emptyset\}$. Since $\{P\} \, S \, \{U\}$ and $\{P\} \, S \, \{B\}$ and $\{U\} \wedge \{B\} \, S \, \{W \neq \emptyset\}$, then $\{P\} \, S \, \{W \neq \emptyset\}$.

Using the Hoare's rule of composition let us conjunct the results of the Step 4:
$\{P\} \, S \, \{D\}$ and $\{P\} \, S \, \{W \neq \emptyset\}$.
Then we get $\{P\} \, S \, \{V\}$, where $V = \{D \vee \{W \neq \emptyset\}\}, S = S1, S2, \ldots, 36$, which means the Theorem 1 is proven.